\documentclass[letter]{IEEEtran}
\usepackage{amsmath,amssymb,graphicx}
\usepackage{bbm}  
\usepackage{physics} 
\usepackage{color,xcolor}
\usepackage{subfigure}
\usepackage{psfrag}
\usepackage{float}
\usepackage{cite}
\usepackage{booktabs}
\usepackage{amsmath}
\usepackage[justification=centering]{caption}
\usepackage{multirow,multicol}
\usepackage{epsfig}
\usepackage{epstopdf}
\usepackage{balance}
\usepackage{mathtools, cuted}
\usepackage{lipsum}
\usepackage{longtable}
\usepackage{array}
\usepackage{booktabs}
\usepackage{tabularx}
\usepackage{booktabs}
\usepackage{enumerate}
\usepackage{graphicx,cite}
\usepackage{times}
\usepackage{amssymb,lipsum}
\usepackage{mathtools}

\usepackage[utf8]{inputenc}
\usepackage[T1]{fontenc}
\usepackage[english]{babel}

\usepackage{amsmath,amssymb}                         
\usepackage[ruled,vlined,linesnumbered]{algorithm2e} 

\begin{document}
\pagenumbering{gobble}
\title{FLARE: Flying Learning Agents for Resource Efficiency in Next-Gen UAV Networks}
\author{
\IEEEauthorblockN{Xuli Cai, Poonam Lohan, and Burak Kantarci}
\IEEEauthorblockA{\textit{University of Ottawa, Ottawa, ON, Canada}\\
\{xcai049, ppoonam, burak.kantarci\}@uottawa.ca}
\vspace{-0.3in}

}
\maketitle

\begin{abstract}
This letter addresses a critical challenge in the context of 6G and beyond wireless networks, the joint optimization of power and bandwidth resource allocation for aerial intelligent platforms, specifically uncrewed aerial vehicles (UAVs), operating in highly dynamic environments with mobile ground user equipment (UEs). We introduce FLARE (Flying Learning Agents for Resource Efficiency), a learning-enabled aerial intelligence framework that jointly optimizes UAV positioning, altitude, transmit power, and bandwidth allocation in real-time. To adapt to UE mobility, we employ Silhouette-based K-Means clustering, enabling dynamic grouping of users and  UAVs' deployment at cluster centroids for efficient service delivery. The problem is modeled as a multi-agent control task, with bandwidth discretized into resource blocks and power treated as a continuous variable. To solve this, our proposed framework, FLARE, employs a hybrid reinforcement learning strategy that combines Multi-Agent Deep Deterministic Policy Gradient (MADDPG) and Deep Q-Network (DQN) to enhance learning efficiency.  Simulation results demonstrate that our method significantly enhances user coverage, achieving a 73.45\% improvement in the number of served users under a 5 Mbps data rate constraint, outperforming MADDPG baseline. 
\end{abstract}


\section{Introduction and Related Work}

Uncrewed Aerial Vehicles (UAVs) are increasingly utilized in next-generation wireless communication systems due to their inherent mobility, altitude control, and rapid deployment capabilities. These features make UAVs well-suited to provide on-demand connectivity for mobile and heterogeneous user equipment (UE) with dynamic service requirements \cite{parvaresh2022tutorial}. However, effective UAV placement, along with joint power and bandwidth (BW) allocation to improve user coverage, remains a critical challenge due to interference and fading effects.

To address these complexities, recent studies have explored the use of Multi-Agent Deep Reinforcement Learning (MADRL), particularly the MADDPG algorithm, for optimizing UAV-enabled communication networks. For instance, the MADDPG-M\&L (MADDPG based on Matching Game and Lagrangian Dual) approach proposed in~\cite{10938906} investigates UAV-assisted user association and slicing resource allocation in heterogeneous networks. In the context of Mobile Edge Computing (MEC), the authors in~\cite{du2024maddpgjointservice} address the problem of joint service placement and task offloading in air–ground integrated networks (AGINs). It uniquely integrates placement and offloading control, including service instance replacement and offload decision coupling.
Other related works, such as ~\cite{10716389} and ~\cite{10673993} leverage MADDPG for UAV dynamic position planning and task offloading. UAV utility maximization strategy proposed in ~\cite{10811139} considers
DQN-DDPG joint BW-power optimization.
 The greedy reinforcement learning (RL) method in ~\cite{kaleem2022enhanced} improves fairness in emergency scenarios. The study in ~\cite{gemayel2025network} addresses network resource optimization for UAV condition monitoring by proposing a machine learning (ML)-driven vibration analysis framework that balances diagnostic accuracy with communication efficiency through data aggregation and dimensionality reduction. UAV placement and backhaul optimization are explored in ~\cite{pham2021joint}, while ~\cite{nguyen2021resource} presents a joint trajectory and resource allocation framework incorporating admission control under fixed user data demands, aiming to maximize the number of admitted users through iterative optimization with soft admission variables. In ~\cite{samir2019trajectory}, joint trajectory and resource allocation for UAVs serving mobile users in vehicular networks is addressed using problem decomposition and iterative methods.

However, these works do not jointly consider UAV and user mobility, along with power and BW resource constraints to enhance user coverage. In this letter, we propose FLARE, a mobility-aware framework that jointly optimizes UAV positioning, altitude, and resource allocation through clustering and a hybrid MADRL approach to maximize the number of served users. The key contributions of this work are as follows:

\begin{itemize}
\item Integration of the Spatio-Temporal Parametric Stepping (STEP) model to capture continuous UE mobility across time frames.
\item Development of a Silhouette-based K-Means clustering technique for adaptive UAV-UE association under dynamic mobility patterns.
\item Design of a hybrid MADDPG+DQN framework with mixed activation functions to enable continuous control of UAV altitude and power allocation, alongside discrete control of BW resource allocation.
\end{itemize}

The remainder of this letter is organized as follows. Section II presents the system model and formulates the joint UAV positioning and resource allocation problem for user coverage maximization. Section III describes the proposed hybrid MADRL-assisted FLARE framework. Section IV presents the numerical results, and Section V concludes the letter.

\section{System Model and Problem Formulation}

\subsection{System Model}
\begin{figure}[t!]
    \centering 
\includegraphics[width=9 cm]{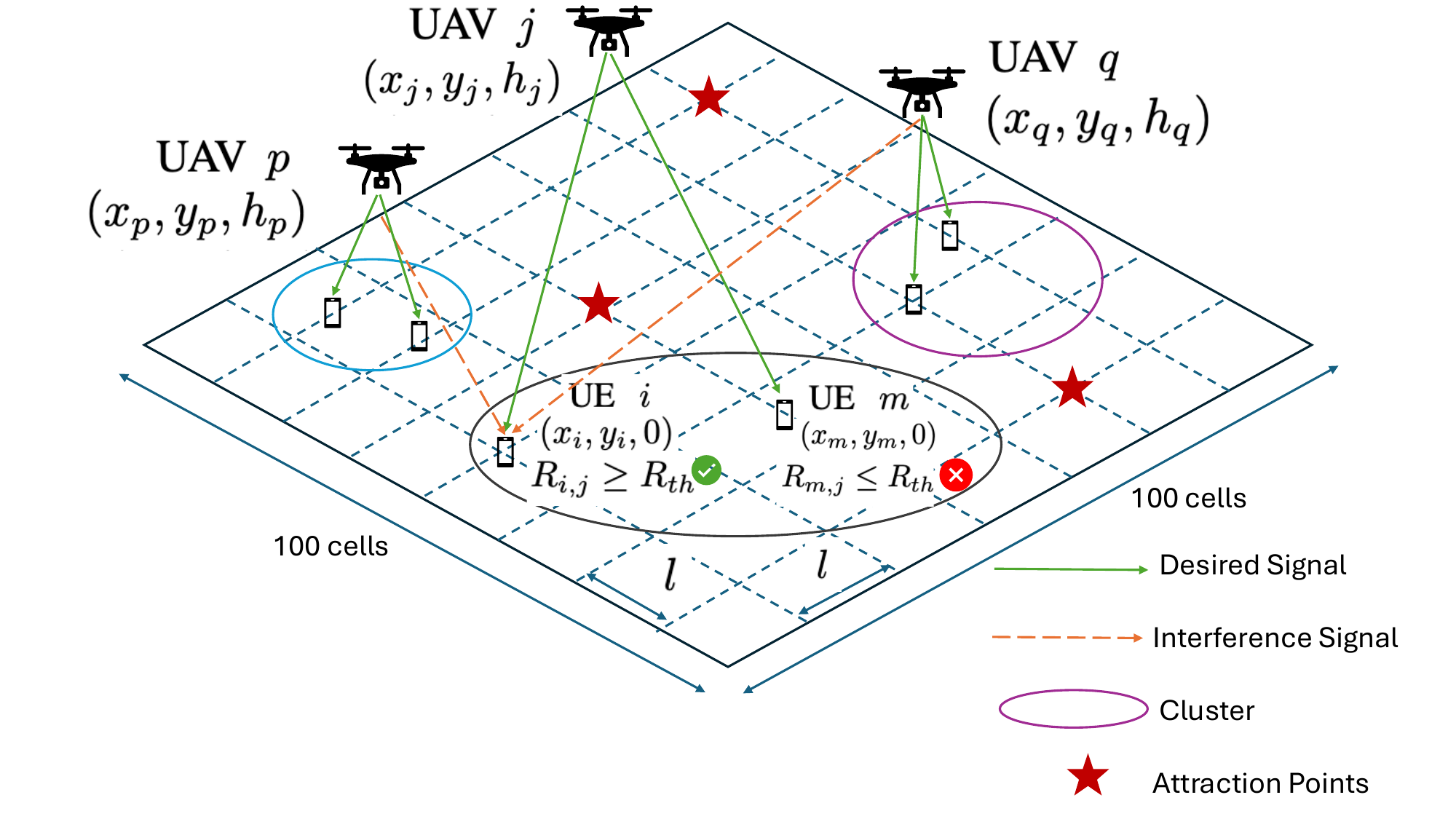}\vspace{-0.5cm}
    \caption{System Diagram}
    \label{fig:problem}
\end{figure}

 We consider a multi-UAV-assisted communication system, where $N$ ground UEs are initially distributed uniformly across a two-dimensional square area $\Psi$, partitioned into a $100 \times 100$ grid with each cell measuring $l$ meters.  Let $I$ denote the set of UEs and we model UE mobility with the STEP model over $F$ time frames. At each time frame, UEs' positions are updated, K-means clustering (with $K$ chosen via the silhouette score) defines clusters, and UAVs are positioned at the centroids of clusters. The clustering method and UAV deployment strategy are detailed in the subsequent section. As depicted in Fig.~\ref{fig:problem}, the scenario involves multiple UAVs, each serving a dynamically formed cluster of UEs.  

The optimization is carried out frame-by-frame over $F$ frames, with updated UE locations and clustering at each frame \( t \in \{1, 2, \dots, F\} \). Each UE is identified by $i\in I{\triangleq}{1,2,\ldots,N}$, with coordinates $(x_i^t,y_i^t,0)$ representing their positions at frame $t$. Let $\mathcal{J}^t$ denote the set of active UAVs at frame $t$. UAV, $j \in  \mathcal{J}^t$, is located at $(x_j^t,y_j^t,h_j^t)$ in 3D space. Considering a UE $i$ at a horizontal distance $d_{i,j}^t\triangleq\sqrt{(x_j^t-x_i^t)^2+(y_j^t-y_i^t)^2}$ from its associated UAV $j$, the Euclidean distance between them is $r_{i,j}^t=\sqrt{{d_{i,j}^t}^2+{h_j^t}^2}$, and the elevation angle is $\theta_{i,j}^t=\sin^{-1}(h_j^t/r_{i,j}^t)$. All UEs and UAVs use a single antenna.

We adopt an air-to-ground channel model with LoS and NLoS link components \cite{Hourani}. LoS and NLoS links experience Rician and Rayleigh fading, respectively. In NLoS links, path loss exponent $\alpha_{NLoS}$ is higher than $\alpha_{LoS}$ due to increased attenuation from shadowing and reflections. The received power  at UE $i$ for LoS and NLoS links from UAV $j$ is:
\begin{equation}
   \text{P}^t_{\text{LoS}_{i,j}} = P_{i,j}^t g_{i,j} {r_{i,j}^t}^{-\alpha_{LoS}}, \quad  
\end{equation}
\begin{equation}
    \text{P}^t_{\text{NLoS}_{i,j}} = P_{i,j}^t k_{i,j} {r_{i,j}^t}^{-\alpha_{NLoS}}, 
\end{equation}
where $P_{i,j}^t$ is the transmit power allocated from UAV $j$ to UE $i$, $ g_{i,j}$ is the dynamic rician fading factor and $ k_{i,j}$ is the dynamic rayleigh fading factor. The LoS link probability between UE $i$ and UAV $j$ is given as \cite{Hourani}:
\begin{equation}\label{eq:plos}
p_{\text{LoS}_{i,j}}^t=\frac{1}{1+c\exp(-b[(180/\pi)\theta_{i,j}^t-c])},
\end{equation}
with $b$ and $c$ being environment-dependent constants. The effective received power becomes:
\begin{equation}\label{eq:eq1}
    \text{P}^t_{\text{eff}_{i,j}} = p_{\text{LoS}_{i,j}}^t \cdot \text{P}^t_{\text{LoS}_{i,j}} + (1 - p_{\text{LoS}_{i,j}}^t) \cdot \text{P}^t_{\text{NLoS}_{i,j}}.
\end{equation}
Considering inter-UAV interference and using Shannon's capacity formula, the achievable data rate of UE $i$ from UAV $j$ is given by:
\begin{equation}\label{eq:eq2}
R_{i,j}^t= B_{i,j}^t\log_2\left(1+{\text{P}^t_{\text{eff}_{i,j}}}/{(I_{i,j}^t+N_o)}\right),
\end{equation}
where $B_{i,j}^t$ denotes the BW allocated to UE $i$, $N_o$ is the additive white Gaussian noise (AWGN) power and $I_{i,j}^t = \sum_{s \neq j}P_{s,\text{avg}}^t k_{i,s} {r_{i,s}^t}^{-\alpha_{NLoS}}$ is the interference from other UAVs. \textcolor{black}{For tractability and consistent with dense urban settings \cite{Hourani}, we assume that only NLoS links contribute to inter-UAV interference, while noting that LoS interference may dominate in open environments and is left for future extension.} UE $i$ is considered served if $R_{i,j}^t \ge R_{th}$, where $R_{th}$ denotes the target data rate threshold. To quantify the number of served users, we define a binary variable $a_{i,j}^{t}$ as follows:
\begin{equation}
a_{i,j}^{t} = 
\begin{cases}
1, & \text{if UE } i \text{ is served by UAV } j \text{ at frame } t, \\
0, & \text{otherwise.}
\end{cases}
\label{eq:binval}
\end{equation}

\subsection{Problem Formulation}
Our goal is to maximize the number of served UEs over the entire mobility duration by dynamically adjusting UAV positions, altitudes, and resource allocations (power and BW).
Let $N_j^t = \sum_{i\in I} a_{i,j}^t$ denote  the number of UEs served by UAV $j$ at frame \(t\). The optimization problem for frame $t$ is formulated as:
\begin{align}
&(\mathcal{P}):\;\max_{(P_j^t,\, B_j^t,\, h_j^t,\,\mathbf{p}_j^t)} \sum_{j\in\mathcal{J}^t} N_j^{t},\;\\
&\text{s.t.:}\; (C1): R_{i,j}^t \mathbbm{1}(a_{i,j}^{t}=1) \ge R_{th},\quad \forall\, i\in I,\, \forall\, j\in\mathcal{J}^t\nonumber\\
&(C2):\sum_{j\in\mathcal{J}^t} a_{i,j}^{t} \le 1,\quad \forall\, i\in I\nonumber\\
&(C3):\; a_{i,j}^{t} \in \{0,1\},\quad \forall\, i\in I,\, \forall\, j\in\mathcal{J}^t\nonumber\\
&(C4): \sum_{l=1}^{N_{j}^t} P_{l,j}^t \le P_{max}; \quad \forall\, j\in\mathcal{J}^t\nonumber\\
&(C5): \sum_{l=1}^{N_{j}^t} B_{l,j}^t \le B_{max}; \quad \forall\, j\in\mathcal{J}^t\nonumber\\
&(C6): \mathbf{p}_j^t = (x_j^t,y_j^t) \in \Psi, \quad \forall\, j\in\mathcal{J}^t\nonumber\\
&(C7): h_{min} \le h_{j}^t \le h_{max}, \quad \forall\, j\in\mathcal{J}^t.\nonumber
\end{align}
 Constraint (C1) ensures that each served UE meets the required rate. (C2) ensures each UE is served by at most one UAV. (C3) enforces binary assignment. (C4) and (C5) restrict UAV transmission power and BW under certain power and BW budgets. (C6) and (C7) ensure UAVs stay within the field and height range. \textcolor{black}{The joint constraints (C1–C7) are challenging to satisfy due to mobility-driven variations in link budgets, interference coupling, and the hybrid discrete–continuous decision space. Due to the dynamic, non-convex nature of this problem and UE mobility, we propose a joint clustering and hybrid MADRL design to adapt UAV positioning, altitude, and resource allocation over time}.
 
\subsection{STEP: UE Mobility Model}
We adopt the spatio‐temporal prediction (STEP) model \cite{10887343} to govern the mobility behavior of \(N\) UEs, which are initially distributed uniformly over a bounded \(100\times100\) grid \(\Psi\). At each time frame $t=1,\dots,F$, the position of UE \(i\) is represented as 
\begin{equation}
p_i^t=(p_{x_i}^t,p_{y_i}^t)\in\Psi ,\quad 0\le p_{x_i}^t,p_{y_i}^t\le99,
\end{equation}
where UEs are constrained to move only to one of the four adjacent grid cells, which means  \(p_i^{t+1}-p_i^t \in\{(\pm1,0),(0,\pm1)\}.\) Movement decisions are influenced by a predefined set of $K$ attraction points \(\mathcal{A}=\{\mathbf{a}_1,\dots,\mathbf{a}_K\}\subset\Psi\).
At each time frame, UE  selects its nearest attraction point \(\mathbf{a}_K\in\mathcal{A}\), and then identifies the valid move \(\mathcal{M}\), minimizing \(\|\mathbf{a}_k-p_i^{t+1}\|^2\). With a fixed probability \(p\in[0,1]\), UE selects the move \(\mathcal{M}\); otherwise, it randomly chooses one of the four adjacent directions from \(\{(\pm1,0),(0,\pm1)\}\) at uniform probability. To prevent collisions, if the selected cell is already occupied during the same frame, the move is resampled until an unoccupied cell is found. The resulting position is then recorded as \(p_i^{t+1}\). Note that physical coordinates are obtained by scaling the grid position via $(x_i^t,y_i^t)=(p_{x_i}^t\times l,\;p_{y_i}^t \times l)$, where $l$ denotes the cell size of one grid unit.

\subsection{Clustering and UAV Positioning}
\textcolor{black}{At each frame \( t \), UEs are grouped using  Silhouette-based K-Means clustering since it can automatically determine the optimal number of clusters at each time frame under UE mobility. It also provides a lightweight and stable alternative to methods such as DBSCAN or spectral clustering, which require sensitive parameter tuning and incur higher computational cost. Clustering improves feasibility by reducing intra-cluster distances, enhancing link budgets, and lowering inter-cluster interference, which helps satisfy (C1)–(C5).}
The clustering quality is evaluated using the silhouette score:
\begin{equation}
s_i^t = {(b_i^t - a_i^t)}/{\max\{a_i^t, b_i^t\}},
\end{equation}
where \( a_i^t \) and \( b_i^t \) denote the average intra-cluster and nearest inter-cluster distances for UE 
$i$ at frame $t$, respectively. The optimal number of clusters is determined as:
\begin{equation}
{k^t}^* = \arg\max_{k \in \{2, \dots, k_{\text{max}}\}} S_k^t,
\end{equation}
where \( S_k^t \) is the average silhouette score across all UEs for $k$ clusters.
Once the optimal cluster count \( {k^t}^* \) is identified, UEs are partitioned into 
\( {k^t}^* \)
  clusters using the K-means algorithm. Each UAV \(j\) is then positioned at the centroid of its assigned cluster \( C_j^t \), with the horizontal coordinates $(x_j^t,y_j^t)$. The corresponding altitude $h_j^t$ is optimized by Algo.~\ref{alg:hierarchical_maddpg_dqn}, aiming to minimize the average UE–UAV distance while adapting to dynamic spatial distributions. UAVs dynamically track the centroids of their assigned clusters, while those not associated with any cluster enter sleep mode to conserve energy. Let $\mathcal{j}^t$ denote the set of active UAVs at time frame $t$.The number of active UAVs, $|\mathcal{J}^t|={k^t}^*$, adapts at each time frame  $t$ according to the spatial distribution of UEs.

\subsection{Joint Multi-Agent Resource Allocation Strategy}
Following UE clustering and UAV placement at the corresponding cluster centroids, we develop a hybrid MADRL frameowrk, FLARE, detailed in Algo.~\ref{alg:hierarchical_maddpg_dqn}. \textcolor{black}{In this framework, MADDPG optimizes each UAV’s altitude \(\{ h_j^t\}\) (C7) and per-user transmit power \( \{P_{i,j}^t\} \) (C4),  while DQN allocates BW \( \{B_{i,j}^t\} \) (C5) to UEs  at each time frame \(t\). }
The framework aims to maximize the number of UEs served while satisfying power and BW constraints at each time frame $t$.

\begin{algorithm}[ht]
\fontsize{8.7pt}{10pt}\selectfont
\SetAlFnt{\fontsize{8.7pt}{10pt}\selectfont}
\SetAlCapFnt{\fontsize{8.7pt}{10pt}\selectfont}
\SetAlCapNameFnt{\fontsize{8.7pt}{10pt}\selectfont}
\SetAlgoNlRelativeSize{-1}
\setlength{\algomargin}{1.2ex}
\SetAlCapSkip{0.4ex}
\caption{Hybrid MADDPG+DQN framework for UAV altitude control and resource allocation to UEs}
\label{alg:hierarchical_maddpg_dqn}
\KwIn{Set of UAVs and UEs, $\mathcal{J}$ and $I$, respectively, No. of episodes $E$, Timesteps $T$,  $P_{\max}$, $B_{\max}$}
\KwOut{Optimized $(h_j^t, \{P_{i,j}^t\}, \{B_{i,j}^t\})$ for each UAV $j\in \mathcal{J}^t$, and each UE $i\in I$}
\textbf{Preprocessing:}  
At each time frame $t$, update UEs' positions via STEP model, cluster UEs with silhouette‐based K‐means, and assign UAVs to each cluster, placing UAV at their centroids\;

\ForEach{UAV agent $j\in \mathcal{J}^t$}{
  Initialize MADDPG actor $\mu_j$, critic $Q_j$, targets $\mu'_j, Q'_j$, replay buffer $\mathcal{D}_{\mathrm{MADDPG}}$ and local state $s_j$\;  
  \For{episode $e=1$ \KwTo $E$}{
    \For{timestep $t=1$ \KwTo $T$}{
      $h_j^t \gets \tanh\bigl(\mu_j^{(h)}(s_j)\bigr)$\;  
      $\{P_{i,j}^t\} \gets \mathrm{softmax}\bigl(\mu_j^{(p)}(s_j)\bigr)$\;
      
      \ForEach{UE $i\in C_j^t$}{
        Initialize DQN $Q_{i,j}$, target $Q'_{i,j}$ for all $i\in C_j^t$, replay buffer $\mathcal{D}_{\mathrm{DQN}}$ and  state $s^{\mathrm{DQN}}_{i,j}$\;
        $s^{\mathrm{DQN}}_{i,j} \!\gets\! \phi\bigl(s_j,\,h_j^t,\,P_{i,j}^t\bigr)$\;  
        $B_{i,j}^t \!\gets\! \epsilon\text{-greedy}\bigl(Q_{i,j},\,s^{\mathrm{DQN}}_{i,j}\bigr)$\;
        Apply the action $B_{i,j}^t$ to UE $i$ and store the experience for $\mathcal{D}_{\mathrm{DQN}}$
      }
      
      Apply joint action $(h_j^t, \{P_{i,j}^t\}, \{B_{i,j}^t\})$ to environment and get observe reward $r_j$, next MADDPG state $s_j'$ and store $(s_j, h_j^t, \mathbf{P}_j^t, r_j, s_j')$ in $\mathcal{D}_{\mathrm{MADDPG}}$\;  
      
      \If{$t \bmod \text{update\_interval} = 0$}{
        Sample minibatch from $\mathcal{D}_{\mathrm{MADDPG}}$\;  
        Update $\mu_j, Q_j$ via gradient steps;  
        $\mu_j' \leftarrow \tau \mu_j + (1-\tau)\mu_j'$,  
        $Q_j' \leftarrow \tau Q_j + (1-\tau)Q_j'$\;
        
        \ForEach{user $i\in C_j$}{
          Sample minibatch from $\mathcal{D}_{\mathrm{DQN}}$\;  
          Update $Q_{i,j}$ via TD‐loss;  
          Soft‐update $Q_{i,j}' \leftarrow \tau Q_{i,j} + (1-\tau)Q_{i,j}'$\;
        }
      }
      $s_j \leftarrow s_j'$\;
    }
  }
}
\end{algorithm}

\subsubsection{Hybrid Action with Mixed Activation}

\textcolor{black}{To support hybrid actions, the MADDPG actor outputs altitude via \texttt{tanh} and per-user power allocation via \texttt{softmax}, ensuring bounded action and reducing infeasible decisions.} This separation avoids constraint violations without penalty terms. UAVs observe local states based on users grouped in its associated \( C_j^t \), and training uses Gaussian exploration noise.

\subsubsection{MADDPG for UAV altitude and transmit power allocation}

Each UAV acts as an agent with:

\begin{itemize}
    \item \textbf{State:} Altitude \( h_j^t \), power \( \{P_{i,j}^t\} \), BW \( \{B_{i,j}^t\} \), and number of served UEs.
    \item \textbf{Action:} Continuous values for \( h_j^t \in [h_{\min}, h_{\max}] \), \( P_{i,j}^t \in [0, P_{\max}] \).
    \item \textbf{Reward:} Number of UEs achieving target data rate.
\end{itemize}

\subsubsection{DQN for bandwidth resource block allocation}
Each UAV integrates a DQN to determine the minimum BW required for each associated UE to satisfy its data rate requirement. BW is allocated in discrete steps, and once a user is successfully served, its allocation is fixed to enable efficient utilization of the remaining resources.

\begin{itemize}
    \item \textbf{State:} \( \{P_{i,j}^t, B_{i,j}^t\} \) for all associated UEs.
    \item \textbf{Action:} Increment or decrement BW resource block.
    \item \textbf{Reward:} 1 if user meets data rate; else 0.
\end{itemize}

The flow of Algo.~\ref{alg:hierarchical_maddpg_dqn} is explained as follows: Line 1 performs preprocessing; lines 2–3 initialize each UAV’s MADDPG; lines 4–5 enter the episode/timestep loops; lines 6–7 compute altitude and power; lines 8–12 run per-UE DQNs to select BW; line 13 execute actions and record transitions; lines 14–19 update MADDPG and DQNs with soft targets; line 20 advance the state. \textcolor{black}{Across episodes, the framework gradually learns stable policies, while within each timestep, DDPG updates the power allocation and DQN adjusts the BW resource blocks accordingly. In this way, both power and BW are adapted jointly until the optimal number of served users is reached.}

\section{Numerical Results}
\subsection{Simulation Setup}

\begin{figure}[ht]
    \centering
\includegraphics[width=6 cm]{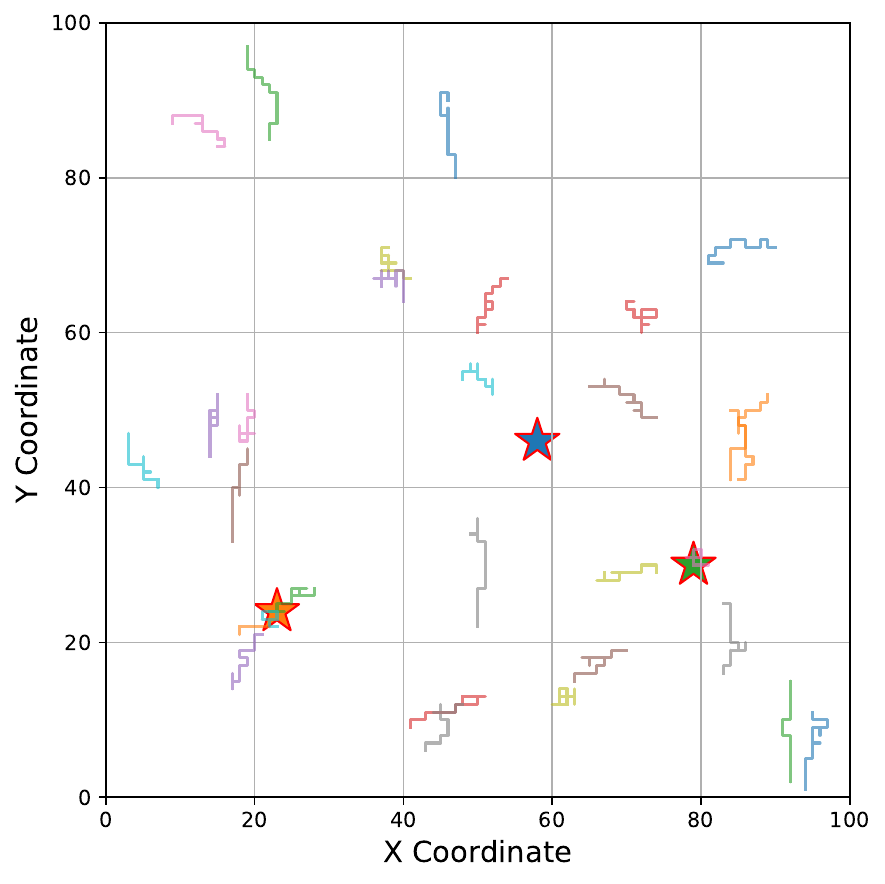}
    \caption{Mobility paths}
    \label{fig:mobility}
\end{figure}

\begin{figure}[H]
    \centering
    \includegraphics[width=7cm]{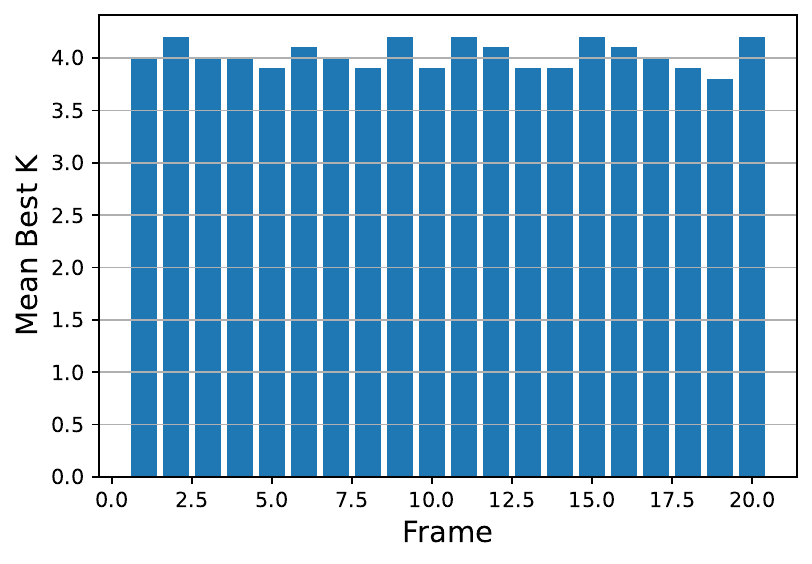}
    \caption{Optimal cluster number based on Silhouette score}
    \label{fig:optimal_cluster}
\end{figure}

We evaluate the proposed hybrid MADRL framework, FLARE, via simulations in a realistic urban environment. Table~\ref{tab:parameters} summarizes the environmental and system settings for a dense urban scenario, while Table~\ref{tab:maddpg_parameters} lists the MADRL training hyperparameters chosen for stable convergence. \textcolor{black}{The lightweight complexity of clustering and sub-millisecond neural inference make the proposed framework practical for real-time UAV deployment.} UE mobility is generated by the STEP model (Fig.~\ref{fig:mobility}), with user trajectories and attraction points; Fig.~\ref{fig:optimal_cluster} plots the mean optimal number of clusters over ten random seeds per frame $t$.  

\begin{table}[ht]
\centering
\caption{Environmental Parameters Used in the Simulations}
\label{tab:parameters}
\begin{tabular}{@{}ccc@{}}
\toprule
\textbf{Symbol}         & \textbf{Description}                      & \textbf{Value} \\ \midrule
\(h_{\min}\), \(h_{\max}\)                   & Min/Max UAV Altitude               & \(300\) m,  \(1000\) m \\
\(l\) & Cell Size &\(300\) m\\ 
\(p\) & Attraction Probability &\(0.4\) \\ 
\(K\) & Number of Attraction Points &\(3\) \\ 
\(N\)                 & Total Number of UEs                    & \(30\)                     \\
\(k_{\max}\)                 & Max Number of Clusters/Agents/UAVs                    & \(5\)  \\
\(P_{\max}\)          & Total Power of each UAV                & \(1\) W                    \\
\(B_{\max}\)                   & BW of each UAV                       & \(3.6\) MHz  \\
\(Block_{size}\)                   & BW size of each block                       & \(18\) kHz  \\
\(Block_{max}\)                   & Block Limit of each UAV                       & \(200\)  \\

\(N_o\)            & Noise Power                              & \(4\times10^{-15}\) W             \\
\(\alpha_{\text{LoS}}\), \(\alpha_{\text{NLoS}}\) & Path Loss Exponent for LoS/NLoS               & \(3\), \(4\)\\
\(c\), \(b\)                   & Environmental Constant \text{(Dense Urban)} & \(11.95\), \(0.136\)         \\
\bottomrule
\end{tabular}
\end{table}

\begin{table}[ht]
\centering
\caption{Model Training Hyperparameters}
\label{tab:maddpg_parameters}
\begin{tabular}{@{}ccc@{}}
\toprule
 \textbf{Description}                           & \textbf{Value} \\ \midrule
 Replay Buffer Size                             & \(100,000\)    \\
 Batch Size,  Update Steps                                     & \(512\) ,  \(2500\)       \\
 Number of Episodes                              & \(100\)         \\
 Steps per Episode                              &  \(500\)         \\
 Actor/Critic Network Learning Rate                    & \(0.0001\)                \\
 Discount Factor for Future Rewards             & \(0.99\)              \\
 Target Network Update Rate                     & \(0.01\)              \\
 Hidden Layer Sizes for Actor and Critic        & \([64, 64]\)        \\
\bottomrule
\end{tabular}
\end{table}

\subsection{Results and Discussion}
\begin{figure}[H]
    \centering
    \includegraphics[width=7 cm]{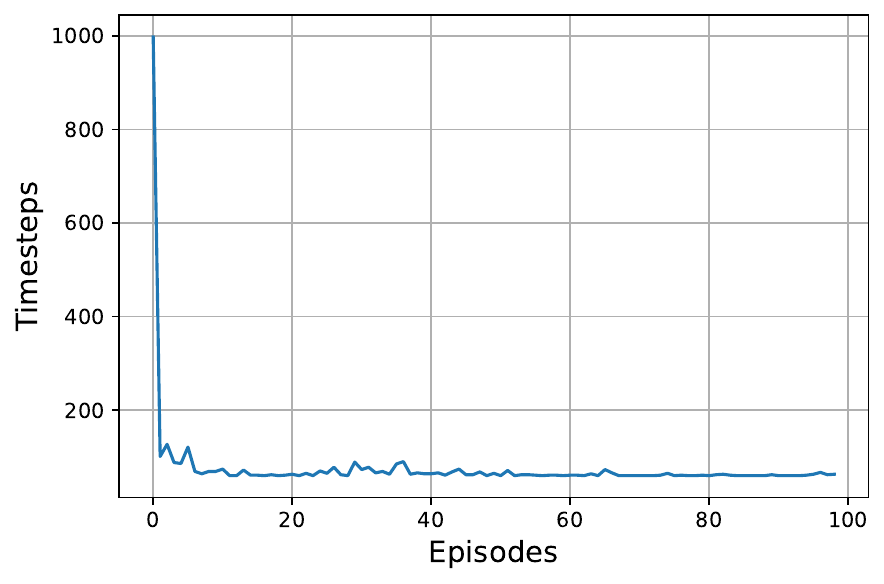}
    \caption{DQN optimal bandwidth searching}
    \label{fig:searching}
\end{figure}

\begin{figure}[H]
    \centering
    \includegraphics[width=7 cm]{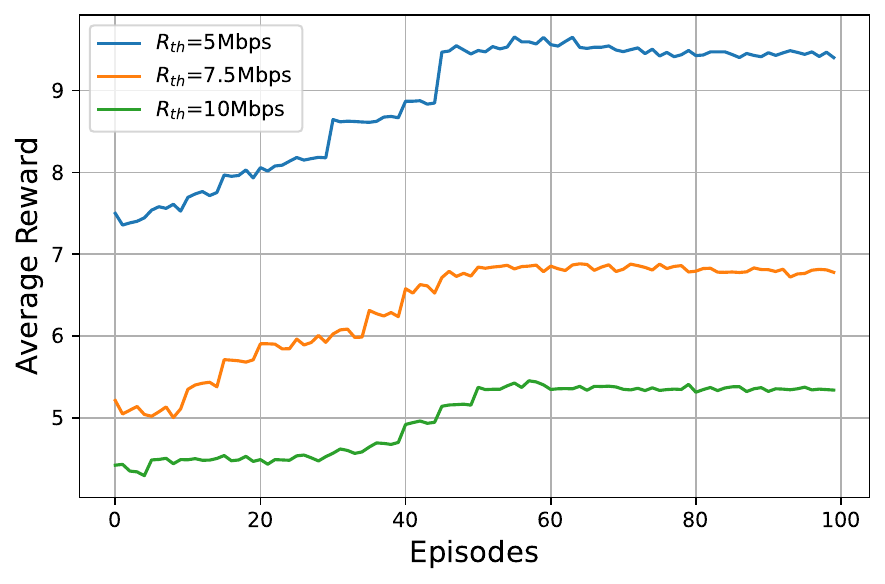}
    \caption{Average reward with training epsiodes}
    \label{fig:convergence}
\end{figure}

Fig.~\ref{fig:searching} illustrates the convergence behavior of the DQN agent during BW block selection. Each episode represents a new attempt to identify the optimal BW allocation, with the objective of minimizing the search time. As training progresses, the agent rapidly converges to optimal solutions, demonstrating improved efficiency over successive episodes.
Fig.~\ref{fig:convergence} demonstrates the stable training behavior of the MADDPG framework under varying data rate thresholds. Following the initial exploration phase using samples from the replay buffer, the average reward increases consistently, indicating effective learning over time.

\begin{figure}[H]
    \centering
    \includegraphics[width=7 cm]{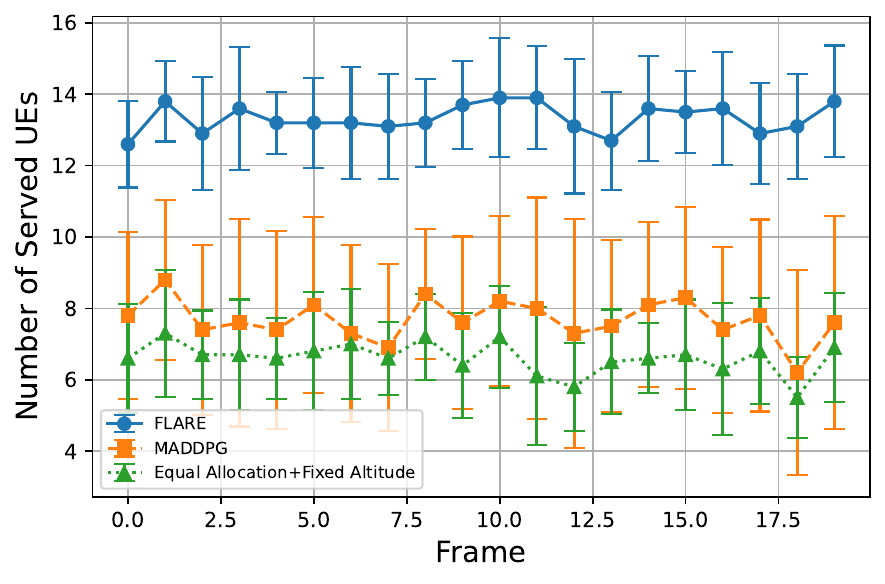}
    \caption{Number of served UEs with $R_{th}=5$Mbps}
    \label{fig:5mbps}
\end{figure}

\begin{figure}[h]
    \centering
    \includegraphics[width=7 cm]{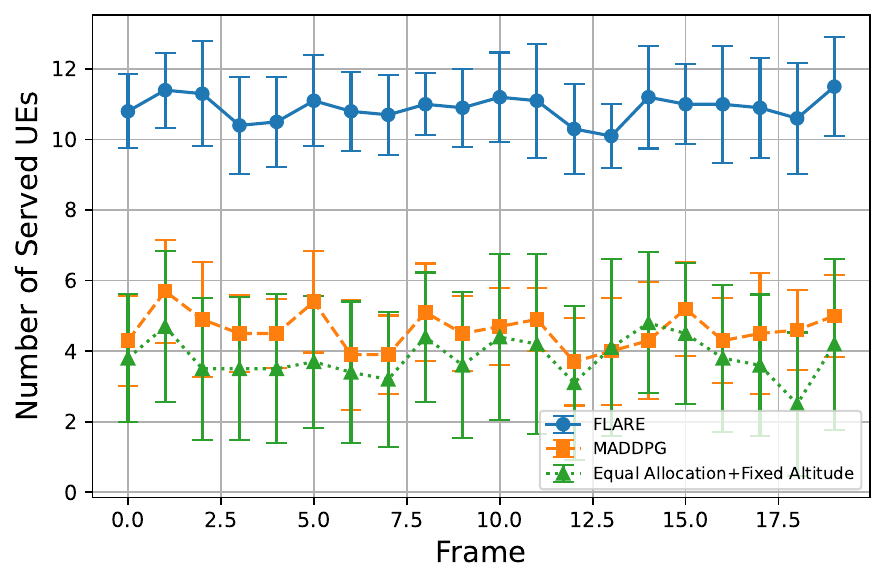}
    \caption{Number of served UEs with $R_{th}=7.5$Mbps}
    \label{fig:7_5mbps}
\end{figure}

Fig.\ref{fig:5mbps} and Fig.\ref{fig:7_5mbps} compare the number of served UEs with target data rates of 5 Mbps and 7.5 Mbps, respectively, under three different methods: our proposed hybrid MADRL framework, FLARE, a baseline MADDPG method, and a static scheme with equal resource allocation and fixed UAV altitude.
Across both scenarios, FLARE demonstrates superior performance by consistently serving a higher number of UEs, enhancing user coverage by 73.45\%  under 5 Mbps and almost 2 times under 7.5 Mbps compared to MADDPG baseline.

\section{Conclusion}

In this letter, we have presented a novel framework, FLARE, to jointly optimize UAV positioning, altitude, power, and discrete bandwidth allocation in dynamic mobile environments for user coverage maximization. Numerical Results confirm that our approach, 
leveraging Silhouette-based K-Mean clustering for adaptive UAV-UE association and a hybrid MADDPG+DQN framework, outperforms the baseline, MADDPG, enhancing user coverage by 73.45\% for 5Mbps data rate threshold. The lightweight complexity of clustering and sub-millisecond neural inference makes the proposed framework practical for real-time UAV deployment. Future work will address security aspects and integrate fairness metrics for a stronger QoS balance with heterogeneous data demands, advancing AI-native aerial access for 6G networks.
\section*{Acknowledgments}
This work is supported in part by the National Science and Engineering Research Council (NSERC) Discovery and NSERC CREATE TRAVERSAL programs.
\bibliographystyle{IEEEtran}

\end{document}